# Locally enhanced conductivity due to tetragonal domain structure in LaAlO$_3$/SrTiO$_3$ heterointerfaces

Beena Kalisky[1,2,†,*], Eric M. Spanton[3,4,†], Hilary Noad[1,4], John R. Kirtley[1], Katja C. Nowack[1,3], Christopher Bell[4], Hiroki K. Sato[4,5], Masayuki Hosoda[4,5], Yanwu Xie[1], Yasuyuki Hikita[4], Carsten Woltmann[6], Georg Pfanzelt[6], Rainer Jany[7], Christoph Richter[7], Harold Y. Hwang[1,4], Jochen Mannhart[6], and Kathryn A. Moler[1,3,4*]

[1]Department of Applied Physics, Stanford University, Stanford, California 94305, USA

[2]Department of Physics, Nano-magnetism Research Center, Institute of Nanotechnology and Advanced Materials, Bar-Ilan University, Ramat-Gan 52900, Israel

[3]Department of Physics, Stanford University, Stanford, California 94305, USA

[4]Stanford Institute for Materials and Energy Sciences, SLAC National Accelerator Laboratory, Menlo Park, California 94025, USA

[5]Department of Advanced Materials Science, The University of Tokyo, Kashiwa, Chiba 277-8561, Japan

[6]Max Planck Institute for Solid State Research, D-70569 Stuttgart, Germany

[7]Experimental Physics VI, Center for Electronic Correlations and Magnetism, Institute of Physics, University of Augsburg, D-86135 Augsburg, Germany.

[†]These authors contributed equally to this work.

The ability to control materials properties through interface engineering is demonstrated by the appearance of conductivity at the interface of certain insulators, most famously the {001} interface of the band insulators LaAlO$_3$ (LAO) and TiO$_2$-terminated SrTiO$_3$ (STO)[1,2]. Transport and other measurements in this system display a plethora of diverse physical phenomena[3-14]. To better understand the interface conductivity, we used scanning superconducting quantum interference device (SQUID) microscopy to image the magnetic field locally generated by current in an interface. At low temperature, we found that the current flowed in highly conductive narrow paths oriented along the crystallographic axes, embedded in a less conductive background. The configuration of these paths changed upon thermal cycling above the STO cubic to tetragonal structural transition temperature, implying that local conductivity is strongly modified by STO tetragonal domain

**structure. The interplay between substrate domains and the interface provides an additional mechanism for understanding and controlling the behaviours of heterostructures.**

The complex oxides display many spectacular phenomena, including colossal magnetoresistance and high-temperature superconductivity that result from their special electronic properties. In recent years, dramatic progress in the difficult task of growing oxide heterostructures has enabled the field of "oxide interface engineering" [15], providing opportunities for electronics. LAO and STO are both band insulators, but when four unit cells or more of LAO are grown on a $TiO_2$-terminated STO {001} substrate, the interface conducts[1,3]. The prevailing explanation of conduction at the interface is electronic reconstruction due to a 'polar catastrophe' in which charge migrates from the top LAO layer to the interface[3,16]. Additional physical phenomena in LAO/STO include superconductivity[3,4,6,7,9,11,17] and magnetism[5,8,10-14]. The in-plane angular dependence of magnetoresistance[18], strong trends in spin-orbit coupling with applied back gate voltage obtained from magnetotransport[9,17,19], and quantum oscillations[9,20] all show a diverse set of behaviours in transport studies, but the possibility of correlation between the transport and the local structure is relatively unexplored. Our previous work [13,14] showed that magnetism and superconductivity in this system can be inhomogeneous. We applied the same local measurement techniques to explore the relationship between structure and normal current flow.

Here we locally probed how current distributed in seven samples (Supplementary Table S1) using a scanning SQUID with micron-scale spatial resolution[21,22]. We applied an AC current to each sample and used lock-in techniques to image the AC flux, which is a convolution of the SQUID's point spread function and the z component of the magnetic field due to local currents (Fig. 1a, Methods).

In a simulated flux image for current flowing in a sample shaped like sample H1 but with uniform conductivity (Fig. 1b), the measured magnetic flux monotonically changes from positive to negative inside the Hall bar, and decays to zero outside. The actual measured magnetic flux in sample H1 (Fig. 1c) similarly varied overall from positive to negative, but substantially deviated from smooth, monotonic behaviour, demonstrating that the current density varies strongly as a function of position.

Although the qualitative conclusion is clear from the raw data, we can better visualize the spatial variations by using an experimentally determined point spread function from a nominally identical SQUID and techniques outlined in Ref. 23 to extract the 2D current density from flux images. The component of the 2D current density parallel to the long dimension of the Hall bar shows a large spatial variation in the magnitude of current flow (Fig. 1d). Line cuts of the current reconstruction reveal that the modulation of the current density is at least a factor of two (Fig. 1e). The height of the actual current modulation and corresponding conductivity modulation may be much higher because the observed variations are resolution-limited (See Supplementary Information).

The current does not flow near the edges of the patterned sample, similar to the behaviour of the superfluid density, which disappears near patterned edges[24].

The paths of higher current density are oriented along the $[100]_p$ STO crystal axis which is also parallel to the edge of the Hall bar (Fig. 1c). The p subscript indicates the pseudocubic axes of the STO substrate. The growth direction is labeled $[001]_p$. Atomic force microscopy on sample H1 revealed that the STO miscut terraces are not parallel to the current paths. This finding excludes one-dimensional electron gas at steps in the interface as the origin of our observations[25].

Flux images of areas of unpatterned sample M1 (Fig. 2) further reveal the structure of the spatial variation in current density. The main features are thin, resolution-limited paths of enhanced current flow, oriented along $[100]_p$ (Fig. 2b), $[010]_p$ (Fig. 2a-c), and $[110]_p$ (Fig. 2d). We did not observe similar features along any other direction. Seven samples were measured and similar features were observed in six of them (Supplementary Table S1). We also observed a small number of features that are not resolution-limited, some of which are aligned along the crystallographic axes (Fig. 2c).

To confirm whether the current density in the narrow paths is higher or lower than the average current density, we considered the geometry of the sample. For the images in Figure 2, we determined the expected direction of current flow based on the position of the contacts and where the image was taken on the sample (Fig. 2). The projection of the current flowing in narrow paths onto the expected current direction is parallel, rather than antiparallel, showing that more current is flowing in the paths. Narrow paths of locally higher current density appear in the middle and bottom panels of Figure 2 as red stripes. Our flux images are dominated by sharp features due to current in narrow paths. Slowly varying fields from a 2D sheet current are more difficult to quantify, so from the present data sets on large samples we cannot accurately determine the fraction of the current that flows in the narrow paths.

The configuration of current paths depends on the thermal history of the sample. We thermally cycled sample M1 repeatedly (Fig. 3a), and AC flux in the sample area was imaged after each cycle at 4 K (Fig. 3b-e). Thermal cycling above a temperature of 105 K, under which the structure of bulk STO transitions from cubic to tetragonal[26], resulted in a new pattern of current paths (compare Fig. 3b to c, d to e); thermal cycling to a temperature below 105 K did not result in a similar change (compare Fig. 3c and 3d). We therefore relate the configuration of the channels to the configuration of STO tetragonal domains.

We imaged sample M1 with polarized light microscopy (Methods), which confirmed the existence of a domain structure that is similar in appearance and behaviour on thermal cycling to current paths observed in these unpatterned samples(Fig. 4, See Supplementary Information)

We checked the dependence of the features on both temperature and back gate voltage. In the unpatterned sample M2, we observed structure in the current flow at ~5 K similar to the features of the type detected in Figures 1 and 2 (See Supplementary Information). The sharp features decayed strongly with temperature, falling beneath the noise level between 40 K and 45 K, indicating that the paths of enhanced conductivity do not exert major influences, if they exist at all, above 40 K. We also investigated normal current flow down to 100 mK in sample H2. Above the superconducting critical temperature, we observed normal current paths oriented along $[110]_p$ and $[1\bar{1}0]_p$. The modulation of the normal current depends on the back gate voltage and disappears below our noise at -100 V (See Supplementary Information).

It is natural to ask how the spatial patterns in the normal current flow relate to superconductivity. Sample H2 was superconducting with no applied gate voltage at our lowest temperature. We measured local diamagnetic susceptibility in sample H2 and detected a small modulation (~1% of the total diamagnetic response) with the same orientation and position as the modulations in the normal current flow (See Supplementary Information).

We now turn to a discussion of possible origins of the enhanced conductivity. Walls between ferroelectric domains are conducting in other perovskite systems that are known to be ferroelectric[27,28], but there is no clear evidence for ferroelectricity in LAO/STO, so we consider different mechanisms.

The thermal history dependence of sample M1 (Fig. 3) indicated that the observed features are due to the tetragonal structure of STO below 105 K. We expect the domain structure near the interface to be determined predominantly by the STO substrate. The tetragonal transition is driven by the rotation of the $TiO_6$ octahedra about the lengthened axis, labeled as the c-axis [26]. The octahedral rotation angle is 2.1° at 4 K, and the sign of the rotation alternates every other octahedron in all three directions[29]. Domains with lengthening along all three original cubic axes can form. The four possible twin planes between two tetragonal domains expressed in terms of the original cubic directions are

$(110)_p$, $(1\bar{1}0)_p$, $(011)_p$, and $(101)_p$ (Fig. 5). Twin boundaries intersect the conducting interface along $[110]_p$, $[1\bar{1}0]_p$, $[100]_p$, and $[010]_p$.

The directions of the narrow paths are consistent with local conductivity enhanced by twin boundaries. Higher conductivity on narrow domains with lengthening along either $[100]_p$ or $[010]_p$, but not both, are also possible scenarios. Higher spatial resolution measurements could distinguish micron-sized domains from domain boundaries. Antiphase boundaries only occur parallel to the primary crystal axes[30] so cannot by themselves explain our observations. A combination of enhanced conduction on boundaries and domains mentioned above may lead to a pattern similar to the one observed.

The presence of LAO will most likely affect the tetragonal domain structure near the interface. The lattice constant of bulk LAO is slightly smaller than that of STO, so the LAO lattice constant is shortened in the growth direction. In addition, more subtle distortions to both the LAO and STO layers have been observed, including lengthening of the $TiO_6$ octahedra in the first layer of STO[31], buckling between the cations and oxygen in the LAO layers,[32] and octahedral distortion and rotation induced by the LAO in the STO layer[33]. All of these effects may be important to the structure of twin boundaries and tetragonal domains near the interface, and therefore useful for understanding the physical interpretation of our results.

Higher conduction due to tetragonal domain structure indicates a local increase in either mobility or carrier density. The mobility of La-doped $SrTiO_3$ has been shown to be strongly enhanced by strain[34], and the strain due to twin boundaries or tetragonal distortion of the interface may similarly decrease the effective mass. We are unaware of any mechanism by which scattering could decrease by a large amount at a twin boundary or in certain domains. The low-energy twin boundaries in STO are not expected to be charged[30], although an increase in carrier density could come from either increased oxygen vacancy density near the boundary or a change in the twin structure near the interface. Detailed theoretical studies of the change of the domain structure near the interface and the electronic details arising from this change are needed to clarify the origin of the locally enhanced conductivity observed.

Factors important for domain formation, such as strain and cooling rate, will also influence the local conductivity, which may manifest itself in transport studies at low temperatures. Anisotropic conductivity and dependence of transport on other direction-dependent tuning parameters, such as an externally applied magnetic field, may also be affected by the presence of enhanced conductivity due to tetragonal domain structure. In addition, the extraction of global physical parameters from transport assuming uniform conduction may have to be modified for samples with a high density of paths of enhanced conduction (e.g. Fig. 1).

The interplay of interfacial conduction and domain structure adds another level of intricacy to the physical phenomena observed in complex oxide heterostructures. This work emphasizes the ability of small structural changes in the perovskite crystal structure to strongly influence the electronic properties of heterostructures. We have shown that the intersection of domain structure and interfaces can exert effects on the physical properties of the materials system, effects that may not be limited to conductivity in other materials. Accounting for and controlling domain structure may enable the engineering of novel and interesting physical systems.

## Methods

Sample H1 was patterned using an $AlO_x$ hard mask. Gold contacts were patterned onto samples M1 and M2 using ion etching and filled by sputtering with ~50 nm of Ti and then ~150 nm of Au. LAO was deposited in partial oxygen pressure using pulsed-laser deposition and were post-annealed in a partial pressure of oxygen. Specific values of temperature and pressure for each sample are provided in Supplementary Information.

We used a scanning SQUID to image magnetic fields from the samples by recording the magnetic flux through a 3-μm pickup loop (Fig. 1a) as a function of position[21,22]. The measured flux is given by $\Phi_s = \int g(x,y)\vec{B} \cdot \vec{da}$ where the integral is taken over the plane of the SQUID, g(x,y) is the point spread function of the pickup loop, and $\vec{B}$ is the magnetic field produced by the sample. We applied an AC current to the sample and used lock-in techniques to image the flux due to local currents in the sample. Thus, each flux image is a convolution of the SQUID pickup loop and the z component of the magnetic field produced by a current. A current-carrying wire appears as a white stripe next to a dark stripe. To provide more intuitive images, we deconvolved the pickup loop's point spread function from the raw data and inverted the resulting magnetic field to produce a current image, as described elsewhere[35]. The RMS magnitude of applied current ranged from 8 μA for the Hall bar to 600 μA for large samples at frequencies between 600 Hz and 5 kHz. We also imaged local magnetic susceptibility using a field coil on the SQUID chip that is concentric with the pickup loop[20].

Measurements on samples H1, M1, and M2 were made using scanning SQUIDs with a base temperature of 4.2 K. Measurements of superconductivity and normal current in sample H2 were made using a scanning SQUID in a dilution refrigerator[21,22].

We used a constant flow cryostat with optical access and polarization preserving components for polarized light microscopy. Light was first linearly polarized and was then reflected off the surface of the sample. The reflected light was analyzed using a second polarizing filter. The contrast in an image was due to spatially dependent rotation of the light's polarization reflected off the sample.

**Acknowledgements**

We thank G.A. Sawatzky, N. Pavlenko, S. Ilani, Y. Yacoby and A. Vailionis for discussions, Y. Yeshurun, E. Zeldov for use of their optical setups, J. Drori, D. Hadad and Y. Shperber for their assistance with the optical measurements and M. E. Huber for assistance in SQUID design and fabrication.

S. Ilani and collaborators have performed complementary measurements by local electrostatic imaging.

This work was primarily supported by the Department of Energy, Office of Basic Energy Sciences, Materials Sciences and Engineering Division, under contract DE-AC02-76SF00515. Y.W.X. acknowledges partial support from the U.S. Air Force Office of Scientific Research (FAQSSO-10-1-0524). J.M. acknowledges financial support by the German Science Foundation (TRR80). B.K. acknowledges support from FENA and the EC grant no. FP7-PEOPLE-2012-CIG-333799.

**Author Contributions** B.K, E.M.S, H.N., J.R.K, and K.C.N. performed the SQUID measurements. B.K. performed polarized-light measurements. B.K., E.M.S, J.R.K. analyzed the data with input from K.A.M. C.B., H.K.S., Y.X., M.H., Y.H. grew samples H1-H5. C.W., G.P., R.J. grew samples M1 and M2. E.M.S, B.K, K.A.M prepared the manuscript with input from all coauthors. H.Y.H, J.M., and K.A.M. guided the work.

**Author Information** The authors declare no competing financial interests. Correspondence and requests for material should be addressed to B.K. (beena@biu.ac.il) and K.A.M. (kmoler@stanford.edu).

**Figure Captions**

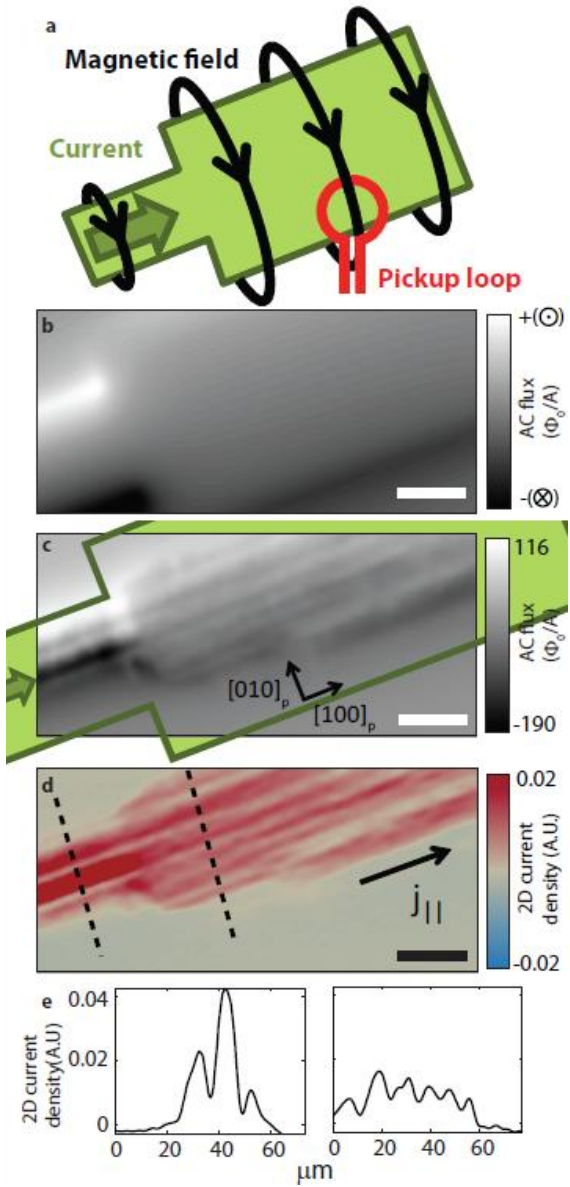

**Figure 1 | Scanning SQUID measurements of current in LAO/STO heterostructures. a,** In our measurement setup, the magnetic flux through the pickup loop (red) from current flowing in the sample is measured as a function of position. **b,** Simulated flux image for the dimensions of sample H1 with uniform conductivity. In AC Flux images, positive flux is flux measured out of the page while negative flux is into the page. **c**, Magnetic flux image of current in patterned sample H1. Green outline indicates the dimensions of the patterned LAO/STO. **d,** Reconstructed current densities along the long dimension of the device are obtained from the raw flux image. **e,** Line cuts through the current densities (dashed lines in **d** reveal a large modulation of the amplitude of the current density with position. Scale bars are 30 μm.

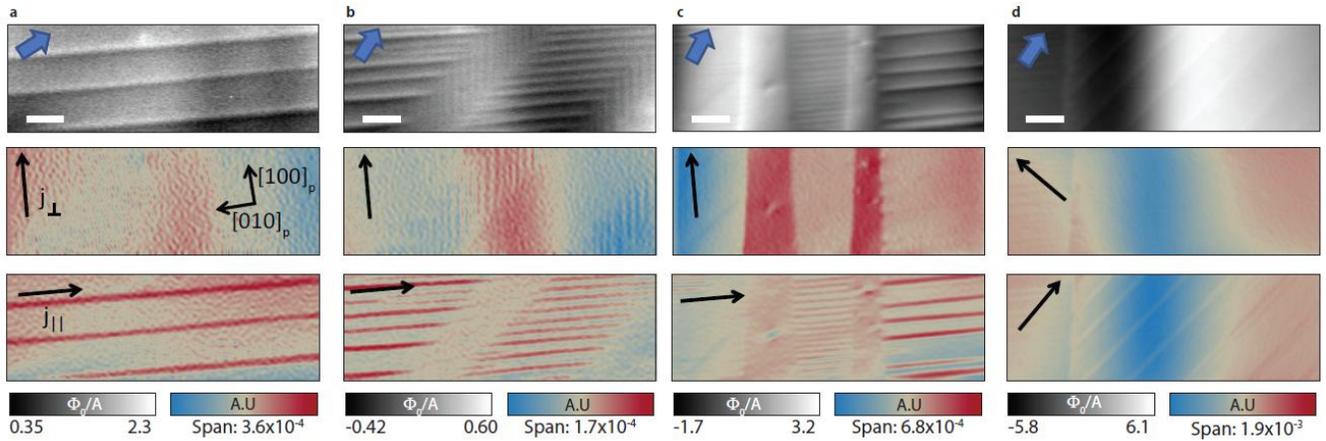

**Figure 2 | Current in unpatterned LAO/STO flows in narrow paths. a-d,** (Top) Magnetic flux images from current flowing through four 250 μm x 85 μm regions of a 5x5 mm sample. The dark and light lines next to each other correspond to a dipolar feature expected for the z component of a magnetic field from a wire-like current. Blue arrows indicate the expected direction of current flowing in a homogeneous sample, showing that more current is flowing in the narrow paths. (Middle, Bottom) Variation in the 2D current density reconstructed from the AC flux parallel (red) and anti-parallel (blue) to the direction indicated with a black arrow. The background current density cannot be determined from the present data and analysis, so red and blue indicate more and less current flowing with respect to an unknown 2D sheet current. All directions of narrow paths observed in sample M1 are shown: **a-c,** $[010]_p$, **b,** $[100]_p$ and **d,** $[110]_p$. Scale bars are 30 μm.

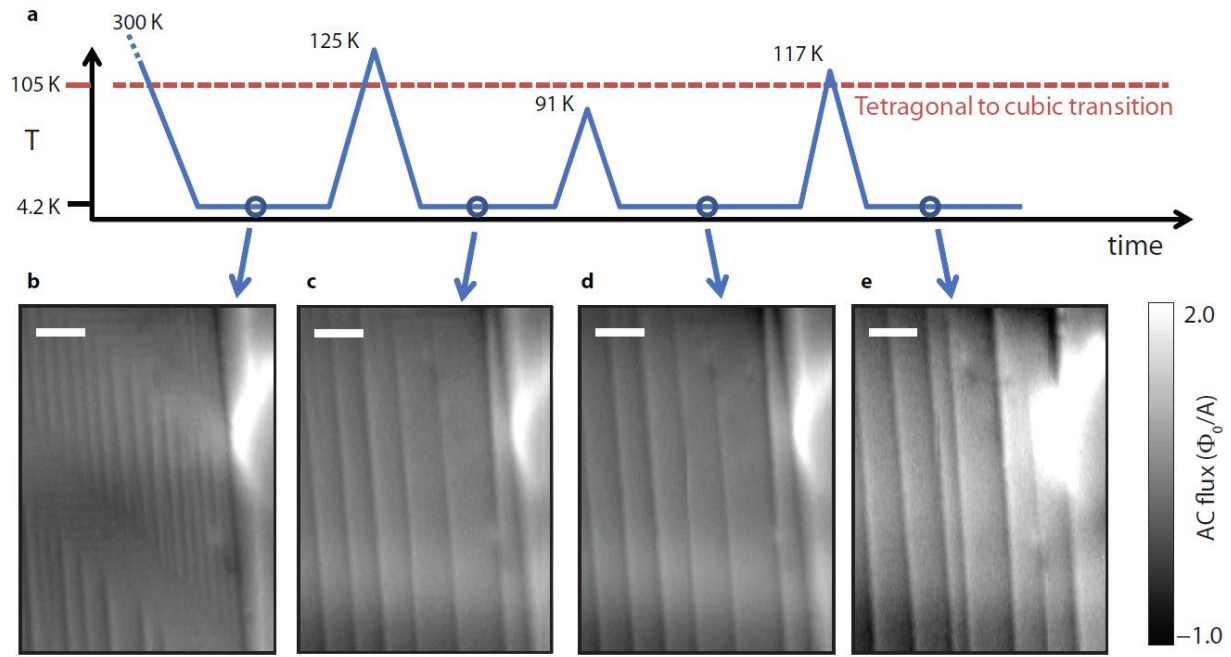

**Figure 3 | Behaviour of unpatterned LAO/STO on thermal cycling. a,** Thermal history of sample M1 with points indicating when the AC flux was imaged. The same area of sample M1 was scanned after initially cooling down from **b,** room temperature and after cycling to **c,** 125 K **d,** 91 K and **e,** 117 K. Reconfiguration of the pattern of AC Flux due to narrow paths of higher current occurred only when the sample was cycled above the STO transition at 105 K, but not below. Scale bars are 30 µm.

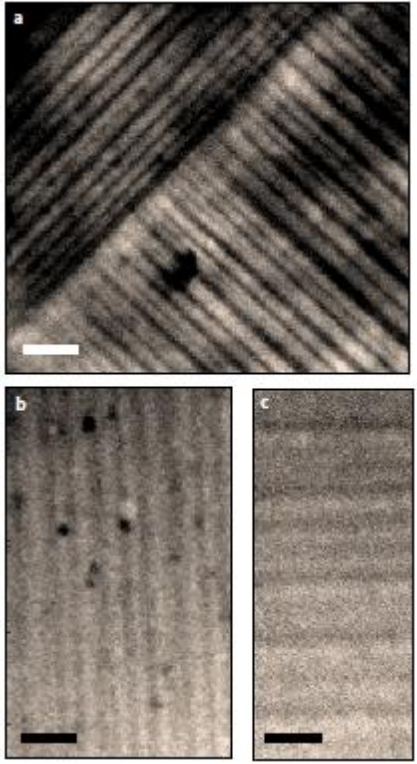

**Figure 4 | Polarized light microscopy reveals the tetragonal domain structure of LAO/STO.**
Polarized light microscopy images of different areas in sample M1 were taken at 20 K **a,** at near-crossed polarizers and **b, c,** at near parallel polarizers. We found domains aligned along the crystallographic directions $[100]_p$, $[010]_p$, $[110]_p$ and $[1\bar{1}0]_p$. Scale bars are 30 μm.

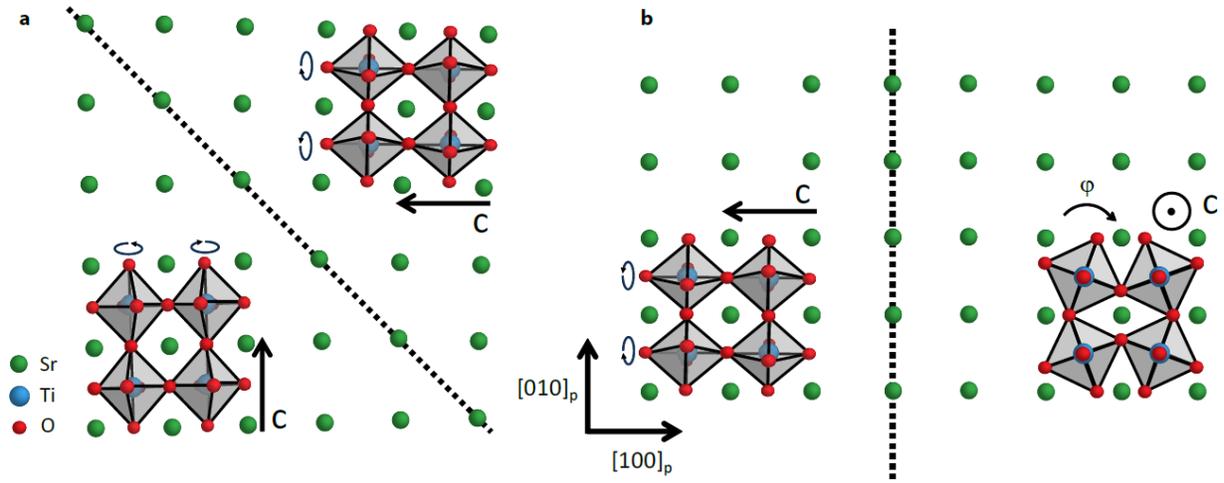

**Figure 5| Schematic twin boundaries and domains in the STO crystal structure in the tetragonal phase**. A view of STO domain structure looking along the growth direction, $[001]_p$. The twin planes **a,** $(110)_p$ and **b,** $(101)_p$ result in four possible geometries (the two shown here and their rotations by 90°). The twin boundary extends in the growth direction in **a** while the boundary shown in **b** extends at approximately 45° to the growth axis. TiO$_6$ octahedral rotations (circular arrows, $\phi$) along the lengthened c-axis (black arrows) are exaggerated, and the exact structure of TiO$_6$ octahedra close to the twin boundary and interface will be more complicated.